\documentclass{article}
\usepackage{arxiv}
\usepackage[utf8]{inputenc} 
\usepackage[T1]{fontenc}    
\usepackage{hyperref}       
\usepackage{url}            
\usepackage{booktabs}       
\usepackage{amsfonts}       
\usepackage{nicefrac}       
\usepackage{microtype}      
\usepackage{lipsum}		
\usepackage{graphicx}
\usepackage{mathrsfs}
\usepackage{doi}
\usepackage{amsmath}
\usepackage{algorithm}
\usepackage{algorithmicx}
\usepackage[noend]{algpseudocode}
\usepackage{caption,subcaption}
\allowdisplaybreaks
\algnewcommand{\LeftComment}[1]{\Statex \(\triangleright\) #1}
\usepackage{stmaryrd}
\usepackage{color,soul}
\usepackage{wrapfig}
\usepackage{cite}
\hypersetup{
    colorlinks=true,
    linkcolor=blue,
    filecolor=magenta,      
    urlcolor=cyan,
}
\usepackage{courier}

\usepackage{listings}
\usepackage{xcolor}
\definecolor{codegreen}{rgb}{0,0.6,0}
\definecolor{codegray}{rgb}{0.5,0.5,0.5}
\definecolor{codepurple}{rgb}{0.58,0,0.82}
\definecolor{backcolour}{rgb}{0.95,0.95,0.92}

\lstdefinestyle{mystyle}{
  backgroundcolor=\color{backcolour},   commentstyle=\color{codegreen},
  keywordstyle=\color{magenta},
  numberstyle=\tiny\color{codegray},
  stringstyle=\color{codepurple},
  basicstyle=\ttfamily\footnotesize,
  breakatwhitespace=false,         
  breaklines=true,                 
  captionpos=b,                    
  keepspaces=true,                 
  numbers=left,                    
  numbersep=5pt,                  
  showspaces=false,                
  showstringspaces=false,
  showtabs=false,                  
  tabsize=2
}

\title{System of equations and staggered solution algorithm for immiscible
two-phase flow coupled with linear poromechanics}

\author{
  {
  Saumik Dana}\\
	University of Southern California\\
	Los Angeles, CA 90007 \\
	\texttt{sdana@usc.edu} \\
}

\date{}

\begin{document}
\maketitle
\begin{abstract}
The governing differential equations of immiscible two phase water-oil flow coupled with linear poromechanics are presented. The solution algorithm poses a constraint on the governing equations of two-phase flow and solves the constrained set of equations using a mixed finite element method. The nonlinear system of equations for the flow model are linearized before solving them in successive Newton-Raphson iterations. The algorithm then solves the governing equations of linear poromechanics using a conforming Galerkin method. The pressure solution from two-phase flow acts as a forcing term for the poromechanics system of equations while the volumetric strain provides poromechanical feedback to the two-phase flow system. 
\end{abstract}
\section{Introduction}
Traditional descriptions of water (or oil) flow in unsaturated deformable porous media neglect the flow of the air phase by assuming that air always remains at atmospheric pressure and that air is displaced without viscous resistance and is free to escape from or to enter into the porous medium. However, flow in such porous media is basically a two phase flow problem. The reader is refered to \cite{liakopoulos} for an excellent rendition of the issues arising out of flow of water through unsaturated porous medium (in the presence of air phase). The reader is also refered to \cite{narasimhannumerical,bernard1,bernard,solidliquidair,khoei,correa} for some of the renditions of numerics for multiphase flow in the presence of deformable porous medium. Now, given that water and oil are liquids whereas air is a gas, the underlying physics of the interaction between solid-water-oil phases is different than that of solid-water-air phases. The first point of departure is the fact that water and oil are both slightly compressible fluids whereas the air is highly compressible. As a result, density variations in case of simultaneous water and air flow are expected to be more significant in comparison to the case of simultaneous water and oil flow. When two phase flow occurs in the presence of deformable porous medium, the poromechanical feedback to flow talks to both the fluid densities and phase saturations in the equations of continuity. As a result, simultaneous water and air flow is expected to be more sensitive than simultaneous water and oil flow to the poromechanical feedback. In lieu of that, the coupling of immiscible water-oil flow with linear poromechanics is the focus of this work. This document is structured as follows: the model equations and staggered solution algorithm are presented in Section 2, the mixed finite element formulation for the flow model is presented in Section 3, the conforming galerking formulation for poromechanics is presented in Section 4 and the conclusions and outlook are presented in Section 5. 
\section{Model equations and algorithm}
\subsection{Flow model}
Let $\Omega \subset \mathbb{R}^3$ be the flow domain with boundary $\partial \Omega=\Gamma_D^f\, \cup \,\Gamma_N^f$ where $\Gamma_D^f$ is Dirichlet boundary and $\Gamma_N^f$ is Neumann boundary. The equations of flow model are
\begin{align*}
\frac{\partial (\phi^* \rho_{\beta}S_{\beta})}{\partial t}+\nabla \cdot \mathbf{z}_{\beta}-\epsilon\frac{\partial (\phi \rho_{\beta}S_{\beta})}{\partial t}=q_{\beta}\qquad (mass\,\,conservation)\\
\mathbf{z}_{\beta}=-\mathbf{K}\lambda_{\beta}(\nabla p_{\beta}-\rho_{\beta} \mathbf{g})\qquad (Darcy's\,\,law)\\
\rho_{\beta}=\rho_{\beta_0}e^{c_{\beta}(p_{\beta}-p_{\beta_0})}\qquad(slightly\,\,compressible\,\,fluid)\\
p_c=p_o-p_w\qquad(capillary\,\,pressure)\\
S_w+S_o=1\qquad(constraint)\\
\mathbf{z}_{\beta}\cdot\mathbf{n}=0 \,\, \mathrm{on}\,\,\Gamma_N^f \times (0,T]\\
p_{\beta}(\mathbf{x},0)=p_{\beta_0}(\mathbf{x}),\,\, \phi(\mathbf{x},0)=\phi_0(\mathbf{x})\qquad \forall \mathbf{x}\in \Omega
\end{align*}
where $\beta\equiv w$ for the water (wetting) phase, $\beta\equiv o$ for the oil (non-wetting) phase, $p_{\beta}:\Omega \times (0,T]\rightarrow \mathbb{R}$ are the phase pressures, $\mathbf{z}_{\beta}:\Omega \times (0,T]\rightarrow \mathbb{R}^3$ are the phase fluxes, $S_{\beta}$ are the phase saturations, $\rho_{\beta}$ are the phase densities, $\phi^*=\phi(1+\epsilon)$ where $\phi$ is the porosity, and $\epsilon$ is the volumetric strain, $\mathbf{n}$ is the unit outward normal on $\Gamma_N^f$, $q_{\beta}$ is the source or sink term, $\mathbf{K}$ is the uniformly symmetric positive definite absolute permeability tensor, $\lambda_{\beta}=k_{\beta}\rho_{\beta}/\mu_{\beta}$ are the phase mobilities where $k_{\beta}$ are the phase relative permeabilities and $\mu_{\beta}$ are the phase viscosities, $c_{\beta}$ are the fluid compressibilities and $T>0$ is the time interval.
\subsubsection{Capillary pressure}
\begin{figure}[h]
\centering
\includegraphics[scale=0.5]{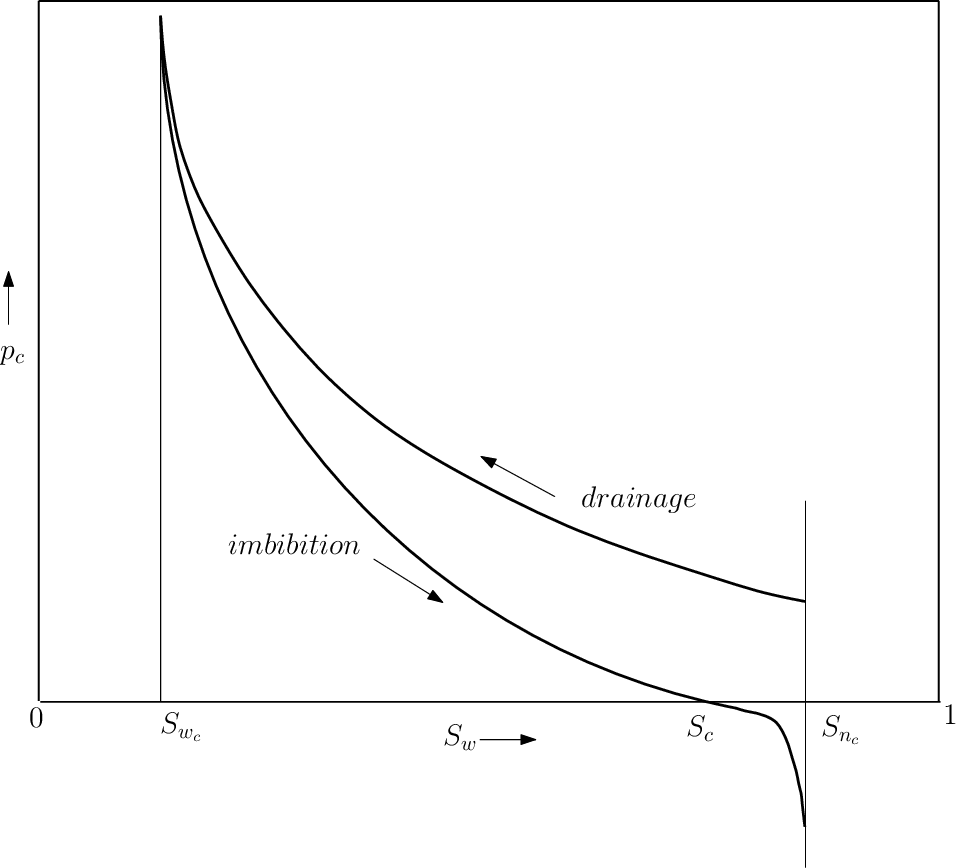}
\caption{Typical capillary pressure curves.}
\label{cappress}
\end{figure}
In two phase flow, a discontinuity in fluid pressure occurs across an interface between any two immiscible fluids (e.g., water and oil). This is a consequence of the interfacial tension
that exists at the interface~\cite{leverett,hassler1944,bear}. The discontinuity between the pressure in the nonwetting phase and that in the wetting phase, is referred to as the capillary
pressure $p_c$.
A typical curve of the capillary pressure is shown in Figure \ref{cappress}. The capillary pressure depends on the wetting phase saturation and the direction of saturation change (drainage or imbibition).
The phenomenon of dependence of the curve on the history of saturation is called hysteresis. The value of water saturation at which the capillary pressure is zero is $S_c$.

\subsubsection{Relative permeability}
\begin{figure}[h]
\centering
\includegraphics[scale=0.5]{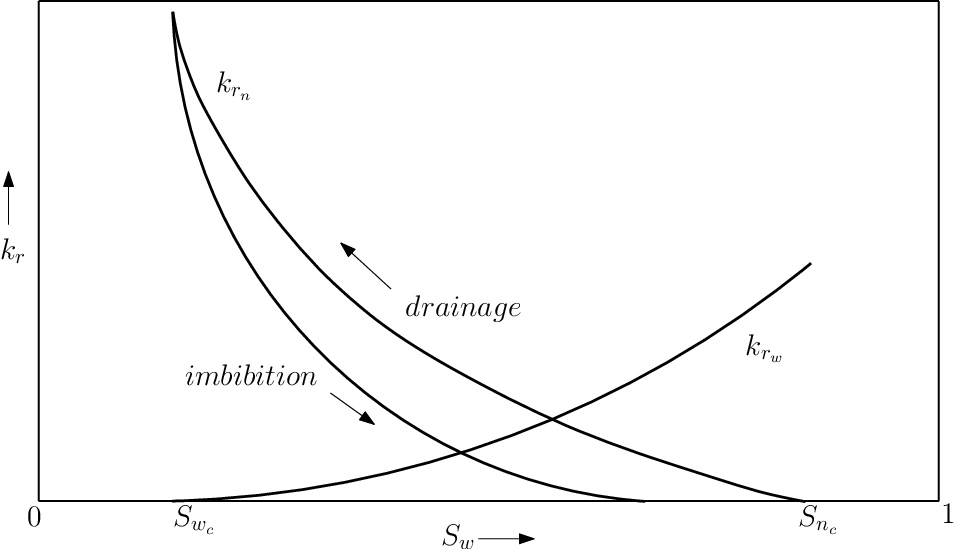}
\caption{Typical relative permeability curves.}
\label{relperm}
\end{figure}
Most of the experimental work on relative permeabilities has been done on two-phase systems~\cite{wyckoff1936,muskat1937,leverett1939,botset,corey,brooks}. Figure \ref{relperm} shows typical results that might be obtained for an oil-water system. As for capillary pressures, relative permeabilities depend not only on the wetting phase
saturation, but also on the direction of saturation change (drainage or imbibition). Note that the curve of imbibition is always lower than that of drainage. For the wetting phase, the relative permeability does not depend on the history of saturation. During the imbibition cycle, the value of $S_w$ at which water starts to flow is called the critical saturation and the value of $S_w$ at which oil ceases to flow is called the residual saturation. Similarly, during the drainage cycle, the value of $S_w$ at which oil starts to flow is called the critical saturation and the value of $S_w$ at which water ceases to flow is called the residual saturation.
In order of avoid the inversion of zero relative permeabilities at residual saturations, the variables $\tilde{\mathbf{z}}_{\beta}=\frac{1}{\lambda_{\beta}}\mathbf{z}_{\beta}$ are introduced~\cite{singhapp} and the Darcy's law is rewritten as $\mathbf{K}^{-1}\tilde{\mathbf{z}}_{\beta}=-\nabla p_{\beta}+\rho_{\beta} \mathbf{g}$.
\subsection{Poromechanics model}
Let $\Omega \subset \mathbb{R}^3$ be the poromechanics domain with boundary $\partial \Omega=\Gamma_D^p \cup \Gamma_N^p$ where $\Gamma_D^p$ is Dirichlet boundary and $\Gamma_N^p$ is Neumann boundary. The equations are
\begin{align*}
\nabla\cdot \boldsymbol{\sigma}+\mathbf{f}=\mathbf{0}\qquad(linear\,\,momentum\,\,balance)\\
\boldsymbol{\sigma}=\mathbb{D}\boldsymbol{\epsilon}-\alpha \bar{p}\mathbf{I}\qquad(constitutive\,\,law)\\
\mathbf{f}=\bar{\rho} \phi\mathbf{g} + \rho_r(1-\phi)\mathbf{g}\\
\bar{p}=p_wS_w+p_oS_o=p_o-p_cS_w\qquad(average\,\,phase\,\,pressure)\\
\bar{\rho}=\rho_wS_w+\rho_oS_o\qquad(average\,\,phase\,\,density)\\
\boldsymbol{\epsilon}(\mathbf{u})=\frac{1}{2}(\nabla \mathbf{u} + \nabla^T \mathbf{u})\\
\mathbf{u}\cdot\mathbf{n}_1=0\,\, \mathrm{on}\,\,\Gamma_D^p \times [0,T],\,\,\boldsymbol{\sigma}^T\mathbf{n}_2=\mathbf{t}\,\,\mathrm{on}\,\,\Gamma_N^p \times [0,T]\\
\mathbf{u}(\mathbf{x},0)=\mathbf{0}\qquad \forall\,\,\mathbf{x}\in \Omega
\end{align*}
where $\mathbf{u}:\Omega \times [0,T]\rightarrow \mathbb{R}^3$ is the solid displacement, $\rho_r$ is the rock density, $\alpha$ is the Biot parameter, $\mathbf{f}$ is body force per unit volume, $\mathbf{t}$ is the traction boundary condition, $\boldsymbol{\epsilon}$ is the strain tensor, $\mathbb{D}$ is the fourth order elasticity tensor and $\mathbf{I}$ is the second order identity tensor.
\subsection{Nature of coupling of immiscible two phase flow with linear poromechanics}
The equation of continuity of slightly compressible single phase flow in non-deformable porous medium (when poromechanics is not active) is parabolic with respect to the pressure variable and degenerates to an elliptic equation when the fluid is assumed incompressible (see \ref{dis1}). As a result, the density variations are expected to be smooth (without any localization) throughout the domain. On the other hand, the equations of continuity in case of immiscible two phase flow in non-deformable porous medium (when poromechanics is not active) degenerate to a hyperbolic equation in one of the phase saturations when capillary pressure is ignored~\cite{azizpetroleum,chavent,chenbook} and fluids are incompressible (see \ref{dis2}). It is well known that in such a scenario, the phase saturations are highly localized in a narrow region of the domain~\cite{buckley,welge,mohsen}. The capillary pressure, when relevant, acts to diffuse the spread of the localization of the phase saturations over a wider domain whereas gravity acts to enhance the localization~\cite{rapoport1953,perkins1957,douglas1958,fayers1959,capillary}. It is clear from the equation of mass conservation that in the presence of deformable porous medium, the volumetric strain computed from the linear momentum balance talks to the fluid densities as well as the phase saturations in the equations of continuity. As a result, the effect of the poromechanical feedback is not only a function of the magnitude of the change in volumetric strain due to the deformation of the porous medium but also the inherent nonlinearities in the flow model.

\subsection{Staggered solution algorithm}
Staggered solution schemes are those in which an operator splitting strategy is used to split the coupled problem into well-posed subproblems which are then solved sequentially instead of solving the coupled system of equations monolithically. The fixed stress split is one such strategy that solves the flow problem while freezing the mean stress followed by the mechanics problem in repeated iterations until a certain convergence criterion is met~\cite{dana-2018,dana2019design,dana2020,dana2021,danacg,danacmame,danathesis}. The mean stress $\sigma_v$ is given as
\begin{align}
\label{expmeanstress}
\sigma_v&=\frac{1}{3}tr(\boldsymbol{\sigma})
=K_b \epsilon-\alpha \bar{p} 
\end{align}
where $K_b$ is the bulk modulus of the solid skeleton or the drained bulk modulus. The basic idea of the fixed - stress split strategy is to solve for the pressure and flux degrees of freedom at the current fixed - stress iteration based on the value of the mean stress from the previous fixed - stress iteration. These pressures then contribute to the force vector in the poromechanics system which is solved for displacements thereby updating the stress state. This updated stress state is then fed back to the flow system for the next fixed - stress iteration. Since this strategy condemns the porous solid to follow a certain stress path during the flow solve, the convergence of the solution algorithm is not automatically guaranteed.
\begin{algorithm}[h]
\begin{algorithmic}
\caption{Algorithm in a nutshell}
\label{stag}
\For{$t \leq T$} \Comment Loop over time steps
\While {$\langle$ convergence criterion not satisfied $\rangle$} \Comment Loop over coupling iterations
\While {$\langle$ convergence criterion not satisfied $\rangle$} \Comment Loop over Newton iterations
\State Solve two phase flow model with mean stress fixed
\EndWhile
\State Update pressures, fluxes and phase saturations
\State Solve linear poromechanics
\State Update volumetric strains
\EndWhile
\EndFor
\end{algorithmic}
\end{algorithm}
\section{Mixed finite element formulation for the flow model}
Let $\mathscr{T}_h$ be the finite element partition of $\Omega$ consisting of elements $E$. The problem statement is : Find $\tilde{\mathbf{z}}_{\beta_h},\mathbf{z}_{\beta_h}\in \mathbf{V}_h$ and $p_{\beta_h},S_w,S_o\in W_h$ such that
\begin{align*}
&\sum\limits_{E\in \mathscr{T}_h}\int\limits_{E}\mathbf{K}^{-1}\tilde{\mathbf{z}}
_{o_h}\cdot \mathbf{v}=\sum\limits_{E\in \mathscr{T}_h}\bigg[\int\limits_{E} p_{o_h}\nabla \cdot \mathbf{v}+\int\limits_{E}\rho_{o} \mathbf{g}\cdot \mathbf{v}\bigg]\\
&\sum\limits_{E\in \mathscr{T}_h}\int\limits_{E}\mathbf{K}^{-1}\tilde{\mathbf{z}}
_{w_h}\cdot \mathbf{v}=\sum\limits_{E\in \mathscr{T}_h}\bigg[\int\limits_{E} p_{o_h}\nabla \cdot \mathbf{v}-\int\limits_{E} p_{c_h}\nabla \cdot \mathbf{v}+\int\limits_{E}\rho_{w} \mathbf{g}\cdot \mathbf{v}\bigg]\\
&\sum\limits_{E\in \mathscr{T}_h}\bigg[\int\limits_{E}\phi^* \rho_{o}S_{o}w-\int\limits_{E}\epsilon\phi \rho_{\beta}S_{o}w+\int\limits_{E}\Delta t\nabla \cdot \mathbf{z}_{o_h}w\bigg]\\
&=\sum\limits_{E\in \mathscr{T}_h}\bigg[\int\limits_{E}(\phi^* \rho_{o}S_{o})^nw-\int\limits_{E}\epsilon(\phi \rho_{o}S_{o})^nw+\int\limits_{E}\Delta tq_{o}w\bigg]\\
&\sum\limits_{E\in \mathscr{T}_h}\bigg[\int\limits_{E}\phi^* \rho_{w}S_{w}w-\int\limits_{E}\epsilon\phi \rho_{w}S_{w}w+\int\limits_{E}\Delta t\nabla \cdot \mathbf{z}_{w_h}w\bigg]\\
&=\sum\limits_{E\in \mathscr{T}_h}\bigg[\int\limits_{E}(\phi^* \rho_{w}S_{w})^nw-\int\limits_{E}\epsilon(\phi \rho_{w}S_{w})^nw+\int\limits_{E}\Delta tq_{w}w\bigg]\\
&\sum\limits_{E\in \mathscr{T}_h}\int\limits_{E}\mathbf{z}_{o_h}
\cdot \mathbf{v}=\sum\limits_{E\in \mathscr{T}_h}\int\limits_{E}\lambda_{o}
\tilde{\mathbf{z}}_{o_h}\cdot \mathbf{v}\\
&\sum\limits_{E\in \mathscr{T}_h}\int\limits_{E}\mathbf{z}_{w_h}
\cdot \mathbf{v}=\sum\limits_{E\in \mathscr{T}_h}\int\limits_{E}\lambda_{w}
\tilde{\mathbf{z}}_{w_h}\cdot \mathbf{v}\\
&\sum\limits_{E\in \mathscr{T}_h}\bigg[\int\limits_{E}S_w w + \int\limits_{E}S_o w\bigg]= \sum\limits_{E\in \mathscr{T}_h}\int\limits_{E}w
\end{align*}
where the terms $(\cdot)^n$ are evaluated at the previous time step, the finite dimensional spaces $W_h$ and $\mathbf{V}_h$ are given as
\begin{align*}
&W_h\equiv \big\{w:w\vert_{E}\in \mathbb{P}_0(E)\,\,\forall E\in \mathscr{T}_h\big\}\\
&\mathbf{V}_h\equiv \big\{\mathbf{v}:\mathbf{v}\vert_{E}\leftrightarrow \hat{\mathbf{v}}\vert_{\hat{E}}:\hat{\mathbf{v}}\vert_{\hat{E}}\in \hat{\mathbf{V}}(\hat{E})\,\,\forall E\in \mathscr{T}_h,\,\,\mathbf{v} \cdot \mathbf{n}=0\,\,\mathrm{on}\,\,\Gamma_N^f\big\}
\end{align*} 
and the details of $\mathbf{\hat{V}}(\hat{E})$ are given in \ref{bddfa}. The nonlinear system of equations is solved using the Newton-Raphson method. Using the notations $(\cdot)^{k}$ for quantity $(\cdot)$ evaluated at the $k^{th}$ Newton iteration and $\delta^{(k)}(\cdot)$ for the change in quantity over the $(k+1)^{th}$ Newton iteration, we arrive at the following
\begin{align}
\label{linearize1}
&\sum\limits_{E\in \mathscr{T}_h}\bigg[\int\limits_{E}\mathbf{K}^{-1}
\delta^{(k)}\tilde{\mathbf{z}}
_{o_h}\cdot \mathbf{v}-\int\limits_{E} \delta^{(k)}p_{o_h}\nabla \cdot \mathbf{v}-\int\limits_{E}\delta^{(k)}\rho_{o} \mathbf{g}\cdot \mathbf{v}\bigg]=R_1\\
\label{linearize2}
&\sum\limits_{E\in \mathscr{T}_h}\bigg[\int\limits_{E}\mathbf{K}^{-1}
\delta^{(k)}\tilde{\mathbf{z}}
_{w_h}\cdot \mathbf{v}-\int\limits_{E} \delta^{(k)}p_{o_h}\nabla \cdot \mathbf{v}+\int\limits_{E} \delta^{(k)}p_{c_h}\nabla \cdot \mathbf{v}-\int\limits_{E}\delta^{(k)}\rho_{w} \mathbf{g}\cdot \mathbf{v}\bigg]=R_2\\
\nonumber
&\sum\limits_{E\in \mathscr{T}_h}\bigg[\int\limits_{E}
\delta^{(k)}\phi^{*}\rho_{o}^kS_{o}^kw+\int\limits_{E}\phi^{*^k}
\delta^{(k)}\rho_{o} S_{o}^{k}w
+\int\limits_{E}\phi^{*^k}
\rho_{o}^{k}\delta^{(k)}S_{o}w\bigg]\\
\nonumber
&-\sum\limits_{E\in \mathscr{T}_h}\bigg[\int\limits_{E}
\delta^{(k)}\epsilon\phi^{k}\rho_{o}^{k}S_{o}^{k}w
+\int\limits_{E}
\epsilon^k\delta^{(k)}\phi\rho_{o}^kS_{o}^kw+\int\limits_{E}\epsilon^k\phi^{k}
\delta^{(k)}\rho_{o} S_{o}^{k}w+\int\limits_{E}
\epsilon^k\phi^{k}\rho_{o}^{k}\delta^{(k)}S_{o}w\bigg]\\
\label{linearize3}
&+\sum\limits_{E\in \mathscr{T}_h}\bigg[\int\limits_{E}
\delta^{(k)}\epsilon(\phi\rho_{o}S_{o})^nw+\int\limits_{E}\Delta t\nabla \cdot \delta^{(k)}\mathbf{z}_{o_h}w\bigg]=R_3\\
\nonumber
&\sum\limits_{E\in \mathscr{T}_h}\bigg[\int\limits_{E}
\delta^{(k)}\phi^{*}\rho_{w}^kS_{w}^kw+\int\limits_{E}\phi^{*^k}
\delta^{(k)}\rho_{w} S_{w}^{k}w
+\int\limits_{E}\phi^{*^k}
\rho_{w}^{k}\delta^{(k)}S_{w}w\bigg]\\
\nonumber
&-\sum\limits_{E\in \mathscr{T}_h}\bigg[\int\limits_{E}
\delta^{(k)}\epsilon\phi^{k}\rho_{w}^{k}S_{w}^{k}w
+\int\limits_{E}
\epsilon^k\delta^{(k)}\phi\rho_{w}^kS_{w}^kw+\int\limits_{E}\epsilon^k\phi^{k}
\delta^{(k)}\rho_{w} S_{w}^{k}w+\int\limits_{E}
\epsilon^k\phi^{k}\rho_{w}^{k}\delta^{(k)}S_{w}w\bigg]\\
\label{linearize4}
&+\sum\limits_{E\in \mathscr{T}_h}\bigg[\int\limits_{E}
\delta^{(k)}\epsilon(\phi\rho_{w}S_{w})^nw+\int\limits_{E}\Delta t\nabla \cdot \delta^{(k)}\mathbf{z}_{w_h}w\bigg]=R_4\\
\label{linearize5}
&\sum\limits_{E\in \mathscr{T}_h}\bigg[\int\limits_{E}\delta^{(k)}\mathbf{z}_{o_h}
\cdot \mathbf{v}-\int\limits_{E}\delta^{(k)}\lambda_o
\tilde{\mathbf{z}}_{o_h}^{k}\cdot \mathbf{v}-\int\limits_{E}\lambda_{o}^{k}
\delta^{(k)}\tilde{\mathbf{z}}_{o_h}\cdot \mathbf{v}\bigg]=R_5\\
\label{linearize6}
&\sum\limits_{E\in \mathscr{T}_h}\bigg[\int\limits_{E}\delta^{(k)}\mathbf{z}_{w_h}
\cdot \mathbf{v}-\int\limits_{E}\delta^{(k)}\lambda_w
\tilde{\mathbf{z}}_{w_h}^{k}\cdot \mathbf{v}-\int\limits_{E}\lambda_{w}^{k}
\delta^{(k)}\tilde{\mathbf{z}}_{w_h}\cdot \mathbf{v}\bigg]=R_6\\
\label{linearize7}
&\sum\limits_{E\in \mathscr{T}_h}\bigg[\int\limits_{E}\delta^{(k)}S_{w}w+\int\limits_{E}\delta^{(k)}S_{o}w\bigg]=R_7
\end{align}
where the residuals are given by
\begin{align*}
&R_1=-\sum\limits_{E\in \mathscr{T}_h}\int\limits_{E}\mathbf{K}^{-1}
\tilde{\mathbf{z}}
_{o_h}^{k}\cdot \mathbf{v}+\sum\limits_{E\in \mathscr{T}_h}\int\limits_{E} p_{o_h}^{k}\nabla \cdot \mathbf{v}+\sum\limits_{E\in \mathscr{T}_h}\int\limits_{E}\rho_{o}^{k} \mathbf{g}\cdot \mathbf{v}\\
&R_2=-\sum\limits_{E\in \mathscr{T}_h}\int\limits_{E}\mathbf{K}^{-1}
\tilde{\mathbf{z}}
_{w_h}^{k}\cdot \mathbf{v}+\sum\limits_{E\in \mathscr{T}_h}\int\limits_{E} p_{o_h}^{k}\nabla \cdot \mathbf{v}-\sum\limits_{E\in \mathscr{T}_h}\int\limits_{E} p_{c_h}^{k}\nabla \cdot \mathbf{v}+\sum\limits_{E\in \mathscr{T}_h}\int\limits_{E}\rho_{w}^{k} \mathbf{g}\cdot \mathbf{v}\\
&R_3=-\sum\limits_{E\in \mathscr{T}_h}\int\limits_{E}\phi^{*^k} \rho_{o}^{k}S_{o}^{k}w-\sum\limits_{E\in \mathscr{T}_h}\int\limits_{E}\Delta t\nabla \cdot \mathbf{z}_{o_h}^{k}w+\sum\limits_{E\in \mathscr{T}_h}\int\limits_{E}\epsilon^k\phi^k \rho_{o}^kS_{o}^kw\\
&\qquad +\sum\limits_{E\in \mathscr{T}_h}\int\limits_{E}
(\phi^* \rho_{o}S_{o})^nw-\sum\limits_{E\in \mathscr{T}_h}\int\limits_{E}\epsilon^k(\phi \rho_{o}S_{o})^nw+\sum\limits_{E\in \mathscr{T}_h}\int\limits_{E}\Delta tq_{o}w\\
&R_4=-\sum\limits_{E\in \mathscr{T}_h}\int\limits_{E}\phi^{*^k} \rho_{w}^{k}S_{w}^{k}w-\sum\limits_{E\in \mathscr{T}_h}\int\limits_{E}\Delta t\nabla \cdot \mathbf{z}_{w_h}^{k}w+\sum\limits_{E\in \mathscr{T}_h}\int\limits_{E}\epsilon^k\phi^k \rho_{w}^kS_{w}^kw\\
&\qquad +\sum\limits_{E\in \mathscr{T}_h}\int\limits_{E}
(\phi^*\rho_{w}S_{w})^nw-\sum\limits_{E\in \mathscr{T}_h}\int\limits_{E}\epsilon^k(\phi \rho_{w}S_{w})^nw+\sum\limits_{E\in \mathscr{T}_h}\int\limits_{E}\Delta tq_{w}w\\
&R_5=-\sum\limits_{E\in \mathscr{T}_h}\int\limits_{E}\mathbf{z}_{o_h}^{k}
\cdot \mathbf{v}+\sum\limits_{E\in \mathscr{T}_h}\int\limits_{E}\lambda_{o}^{k}
\tilde{\mathbf{z}}_{o_h}^{k}\cdot \mathbf{v}\\
&R_6=-\sum\limits_{E\in \mathscr{T}_h}\int\limits_{E}\mathbf{z}_{w_h}^{k}
\cdot \mathbf{v}+\sum\limits_{E\in \mathscr{T}_h}\int\limits_{E}\lambda_{w}^{k}
\tilde{\mathbf{z}}_{w_h}^{k}\cdot \mathbf{v}\\
&R_7=-\sum\limits_{E\in \mathscr{T}_h}\int\limits_{E}S_{w}^{k}w-\sum\limits_{E\in \mathscr{T}_h}\int\limits_{E}S_{o}^{k}w+\sum\limits_{E\in \mathscr{T}_h}\int\limits_{E}w
\end{align*}
\subsection{Imposition of fixed mean stress constraint in every Newton iteration}
The Eulerian porosity variation in a deformable porous medium is approximated as~\cite{geertsma,brown}
\begin{align*}
\delta\phi=\big(\frac{\alpha-\phi}{K_b}\big)(\delta\sigma_v+\delta \bar{p})
\end{align*}
Imposing the fixed mean stress constraint $\delta\sigma_v=0$ results in
\begin{align*}
\delta\phi=\big(\frac{\alpha-\phi}{K_b}\big)\delta \bar{p}\implies \frac{\partial \phi}{\partial \bar{p}}=\frac{\alpha-\phi}{K_b}
\end{align*}
Using the relation $\sigma_v=K_b \epsilon-\alpha \bar{p}$ and the constraint $\delta\sigma_v=0$ results in
\begin{align*}
\delta \epsilon=\frac{\alpha}{K_b}\delta \bar{p}\implies \frac{\partial \epsilon}{\partial \bar{p}}=\frac{\alpha}{K_b}
\end{align*}
In lieu of the relationship $\phi^*=\phi(1+\epsilon)$, the Lagrangian porosity variation is given by
\begin{align*}
&\delta \phi^*= \phi \delta \epsilon + (1+\epsilon)\delta \phi=\big((\phi+1+\epsilon) \frac{\alpha}{K_b}-\frac{\phi^*}{K_b}\big)\delta \bar{p}
\\
&\implies \frac{\partial \phi^*}{\partial \bar{p}}=(\phi+1+\epsilon) \frac{\alpha}{K_b}-\frac{\phi^*}{K_b}
\end{align*} 
The variations of fluid fraction, volumetric strain and porosity are
\begin{align*}
&\delta^{(k)}\phi^{*}=\frac{\partial \phi^*}{\partial \bar{p}}\delta^{(k)}\bar{p}_h=((\phi+1+\epsilon) \frac{\alpha}{K_b}-\frac{\phi^*}{K_b})\delta^{(k)}\bar{p}_h\\
&\delta^{(k)}\epsilon=\frac{\partial \epsilon}{\partial \bar{p}}\delta^{(k)}\bar{p}_h=\frac{\alpha}{K_b}\delta^{(k)}\bar{p}_h\\
&\delta^{(k)}\phi=\frac{\partial \phi}{\partial \bar{p}}\delta^{(k)}\bar{p}_h=(\frac{\alpha-\phi}{K_b})\delta^{(k)}\bar{p}_h
\end{align*}
\subsection{Evaluation of auxiliary derivatives}
The variations of average phase pressure, capillary pressure and densities are
\begin{align*}
&\delta^{(k)}\bar{p}_h=\frac{\partial \bar{p}}{\partial p_o}\delta^{(k)}p_{o_h}+\frac{\partial \bar{p}}{\partial S_w}\delta^{(k)}S_{w}=\delta^{(k)}p_{o_h}-(\frac{\partial p_c}{\partial S_w}S_{w}^k+p_{c_h}^k)\delta^{(k)}S_{w}\\
&\delta^{(k)}p_{c_h}=\frac{\partial p_c}{\partial S_w}\delta^{(k)}S_{w}\\
&\delta^{(k)}\rho_{o}=c_o\rho_{o}^{k}\delta^{(k)}p_{o_h}\\
&\delta^{(k)}\rho_{w}=c_w\rho_{w}^{k}(\delta^{(k)}p_{o_h}-\frac{\partial p_c}{\partial S_w}\delta^{(k)}S_{w})
\end{align*}
where the quantity $\frac{\partial p_c}{\partial S_w}$ is obtained from the capillary pressure curve given in Figure \ref{cappress}. The variations of phase mobilities are
\begin{align*}
&\delta^{(k)}\lambda_o=\frac{\partial \lambda_o}{\partial p_o}\delta^{(k)}p_{o_h}+\frac{\partial \lambda_o}{\partial S_w}\delta^{(k)}S_{w}\\
&\delta^{(k)}\lambda_w=\frac{\partial \lambda_w}{\partial p_o}\delta^{(k)}p_{o_h}+\frac{\partial \lambda_w}{\partial S_w}\delta^{(k)}S_{w}
\end{align*}
where the derivatives are obtained as 
\begin{align*}
&\frac{\partial \lambda_o}{\partial p_o}=\frac{\partial (k_o\rho_o/\mu_o)}{\partial p_o}=\frac{c_ok_o\rho_o}{\mu_o}\\
&\frac{\partial \lambda_o}{\partial S_w}=\frac{\partial (k_o\rho_o/\mu_o)}{\partial S_w}=\frac{\rho_o}{\mu_o}\frac{\partial k_o}{\partial S_w}\\
&\frac{\partial \lambda_w}{\partial p_o}=\frac{\partial (k_w\rho_w/\mu_w)}{\partial p_o}=\frac{c_wk_w\rho_w}{\mu_w}\\
&\frac{\partial \lambda_w}{\partial S_w}=\frac{\partial (k_w\rho_w/\mu_w)}{\partial S_w}=\frac{\rho_w}{\mu_w}\frac{\partial k_w}{\partial S_w}
\end{align*}
where the quantities $\frac{\partial k_o}{\partial S_w}$ and $\frac{\partial k_w}{\partial S_w}$ are obtained from the relative permeability curves given in Figure \ref{relperm}.
\section{Continuous galerkin formulation for the poromechanics model}
The problem statement is : find $\mathbf{u}_h\in \mathbf{U}_h$ such that
\begin{align}
\label{pone1}
&\sum\limits_{E\in \mathscr{T}_h}\int\limits_{E} \boldsymbol{\epsilon}(\mathbf{q}):\mathbb{D}\boldsymbol{\epsilon}(\mathbf{u}_h)=
\sum\limits_{E\in \mathscr{T}_h}\int\limits_{E} \alpha \bar{p} \nabla \cdot \mathbf{q}+\sum\limits_{E\in \mathscr{T}_h}\int\limits_{E}\mathbf{q}\cdot \mathbf{f}+
\sum\limits_{E\in \mathscr{T}_h}\int\limits_{\partial E \cap \Gamma_N^p} \mathbf{q}\cdot \mathbf{t}
\end{align} 
where the finite dimensional space $\mathbf{U}_h$ is given by
\begin{align*}
\mathbf{U}_h=\big\{\mathbf{q}=(u,v,w):u\vert_{E},
v\vert_{E},w\vert_{E}\in \mathbb{Q}_1(E)\,\,\forall E\in \mathscr{T}_h,\mathbf{q}=\mathbf{0}\,\,\mathrm{on}\,\,\Gamma_D^p\big\}
\end{align*}
The discrete displacements $\mathbf{u}_h$ and the corresponding strain tensor $\boldsymbol{\epsilon}(\mathbf{u}_h)$ are written in terms of the nodal displacement degrees of freedom represented by $\mathcal{U}$ as
\begin{equation}
\label{ptwo}
\left.\begin{array}{c}
\mathbf{u}_h = \sum\limits_{E\in \mathscr{T}_h}\mathbf{N}\mathcal{U}\\
\boldsymbol{\epsilon}(\mathbf{u}_h) = \sum\limits_{E\in \mathscr{T}_h}\mathbf{B}\mathcal{U}
\end{array}\right\}
\end{equation}
where $\mathbf{N}$ is the shape function matrix and $\mathbf{B}$ is the strain-displacement interpolation matrix. Equations \eqref{pone1} and \eqref{ptwo} eventually lead to the following system of equations
\begin{equation}
\label{poroeq1}
\left.\begin{array}{c}
\mathbf{K}\mathcal{U}=\mathbf{F}\\
\mathbf{K}=\sum\limits_{E\in \mathscr{T}_h}\int\limits_{E} \mathbf{B}^T \mathbb{D} \mathbf{B}\\
\mathbf{F}=\sum\limits_{E\in \mathscr{T}_h}[\int\limits_{E} \mathbf{B}^T \alpha \bar{p} \mathbf{I}+\int\limits_{E} \mathbf{N}^T\mathbf{f}
+\int\limits_{\partial E\cap \Gamma_N^p} \mathbf{N}^T\mathbf{t}]
\end{array}\right\}
\end{equation}
where $\mathbf{K}$ and $\mathbf{F}$ are refered to as the global stiffness matrix and force vector respectively.
To simplify the computations, \eqref{poroeq1} is recast in compact engineering notation~\cite{ref31} wherein stresses $\boldsymbol{\sigma}$, strains $\boldsymbol{\epsilon}$ and identity tensor $\mathbf{I}$ are represented as vectors and fourth order tensor $\mathbb{D}$ is represented as a second order tensor given by
\begin{align*}
\mathbb{D}=\begin{bmatrix}
\lambda+2G & \lambda & \lambda & 0 & 0 & 0\\
\lambda & \lambda+2G & \lambda & 0 & 0 & 0\\
\lambda & \lambda & \lambda+2G & 0 & 0 & 0\\
0 & 0 & 0 & G & 0 & 0\\
0 & 0 & 0 & 0 & G & 0\\
0 & 0 & 0 & 0 & 0 & G
\end{bmatrix}_{6\times 6}
\end{align*}
where $\lambda$ is the Lame parameter corresponding to the drained response of the porous specimen and $G$ is the shear modulus of porous specimen. The matrices $\mathbf{N}$ and $\mathbf{B}$ are also recast appropriately. 
\subsection{Evaluation of integrals for the poromechanical solve using quadrature}\label{quadrature}
\setcounter{MaxMatrixCols}{30}
Let $\mathbf{r}_i$, $i=1,..,8$ be the vertices of $E$ and let $\hat{E}$ represent a reference unit cube with vertices $\hat{\mathbf{r}}_1\equiv (0,0,0)$, $\hat{\mathbf{r}}_2\equiv (1,0,0)$, $\hat{\mathbf{r}}_3\equiv (1,1,0)$, $\hat{\mathbf{r}}_4\equiv (0,1,0)$, $\hat{\mathbf{r}}_5\equiv (0,0,1)$, $\hat{\mathbf{r}}_6\equiv (1,0,1)$, $\hat{\mathbf{r}}_7\equiv (1,1,1)$ and $\hat{\mathbf{r}}_8\equiv (0,1,1)$.
The trilinear function $F_E(\hat{\mathbf{r}}):\hat{\mathbf{r}}\mapsto \mathbf{r}$ can also be written as
\begin{align*}
F_E(\hat{\mathbf{r}}):\hat{\mathbf{r}}\mapsto \mathbf{r}=\sum\limits_{i=1}^8 \hat{N}_i(\hat{\mathbf{r}})\mathbf{r}_i
\end{align*}
where $\hat{N}_i(\hat{\mathbf{r}})$ are refered to as the shape functions and satisfy the property that $\hat{N}_i(\hat{\mathbf{r}}_j)=\delta_{ij}$, $i,j=1,...,8$ where $\delta_{ij}$ is the Kronecker delta, and are given as
\begin{align*}
\hat{N}_1=(1-\hat{x})(1-\hat{y})(1-\hat{z})\\ 
\hat{N}_2=\hat{x}(1-\hat{y})(1-\hat{z})\\
\hat{N}_3=\hat{x}\hat{y}(1-\hat{z})\\
\hat{N}_4=(1-\hat{x})\hat{y}(1-\hat{z})\\
\hat{N}_5=(1-\hat{x})(1-\hat{y})\hat{z}\\
\hat{N}_6=\hat{x}(1-\hat{y})\hat{z}\\
\hat{N}_7=\hat{x}\hat{y}\hat{z}\\
\hat{N}_8=(1-\hat{x})\hat{y}\hat{z}
\end{align*}
The jacobian of $F_E(\hat{\mathbf{r}})$ is the  given as
\begin{align*}
DF_E(\mathbf{\hat{r}})\equiv\begin{bmatrix}
\frac{\partial x}{\partial \hat{x}} & \frac{\partial x}{\partial \hat{y}} & \frac{\partial x}{\partial \hat{z}}\\
\frac{\partial y}{\partial \hat{x}} & \frac{\partial y}{\partial \hat{y}} & \frac{\partial y}{\partial \hat{z}}\\
\frac{\partial z}{\partial \hat{x}} & \frac{\partial z}{\partial \hat{y}} & \frac{\partial z}{\partial \hat{z}}
\end{bmatrix}=\begin{bmatrix}
\sum\limits_{i=1}^8\frac{\partial \hat{N}_i}{\partial \hat{x}}x_i & \sum\limits_{i=1}^8\frac{\partial \hat{N}_i}{\partial \hat{y}}x_i & \sum\limits_{i=1}^8\frac{\partial \hat{N}_i}{\partial \hat{z}}x_i\\
\sum\limits_{i=1}^8\frac{\partial \hat{N}_i}{\partial \hat{x}}y_i & \sum\limits_{i=1}^8\frac{\partial \hat{N}_i}{\partial \hat{y}}y_i & \sum\limits_{i=1}^8\frac{\partial \hat{N}_i}{\partial \hat{z}}y_i\\
\sum\limits_{i=1}^8\frac{\partial \hat{N}_i}{\partial \hat{x}}z_i & \sum\limits_{i=1}^8\frac{\partial \hat{N}_i}{\partial \hat{y}}z_i & \sum\limits_{i=1}^8\frac{\partial \hat{N}_i}{\partial \hat{z}}z_i
\end{bmatrix}
\end{align*}
The shape function matrix is given as
\begin{align*}
&\hat{\mathbf{N}}=
\begin{bmatrix}
\bar{\mathbf{N}}_1 &\bar{\mathbf{N}}_2 &\bar{\mathbf{N}}_3 &\bar{\mathbf{N}}_4 &\bar{\mathbf{N}}_5 &\bar{\mathbf{N}}_6 &\bar{\mathbf{N}}_7 &\bar{\mathbf{N}}_8  
\end{bmatrix}_{3\times 24}\\
\tag{$\forall\,\,i=1,..,8$}
&\bar{\mathbf{N}}_i=\begin{bmatrix}
\hat{N}_i & 0 & 0\\
0 & \hat{N}_i & 0\\
0 & 0 & \hat{N}_i
\end{bmatrix}_{3\times 3}
\end{align*}
\setlength\arraycolsep{5pt}
The strain-displacement interpolation matrix is given as
\begin{align*}
&\hat{\mathbf{B}}=
\begin{bmatrix}
\hat{\mathbf{B}}_1 &\hat{\mathbf{B}}_2 &\hat{\mathbf{B}}_3 &\hat{\mathbf{B}}_4 &\hat{\mathbf{B}}_5 &\hat{\mathbf{B}}_6 &\hat{\mathbf{B}}_7 &\hat{\mathbf{B}}_8  
\end{bmatrix}_{6\times 24}\\
\tag{$\forall\,\,i=1,..,8$}
&\hat{\mathbf{B}}_i=\begin{bmatrix}
\frac{\partial \hat{\mathbf{N}}_i}{\partial x} & 0 & 0\\
0 & \frac{\partial \hat{\mathbf{N}}_i}{\partial y} & 0\\
0 & 0 & \frac{\partial \hat{\mathbf{N}}_i}{\partial z}\\
\frac{\partial \hat{\mathbf{N}}_i}{\partial y} & \frac{\partial \hat{\mathbf{N}}_i}{\partial x} & 0\\
\frac{\partial \hat{\mathbf{N}}_i}{\partial z} & 0 & \frac{\partial \hat{\mathbf{N}}_i}{\partial x}\\
0 & \frac{\partial \hat{\mathbf{N}}_i}{\partial z} & \frac{\partial \hat{\mathbf{N}}_i}{\partial y}
\end{bmatrix}_{6\times 3}
\end{align*}
where the derivatives of the shape functions are obtained using
\begin{equation*}
\left\{\begin{array}{c}
\frac{\partial \hat{\mathbf{N}}_i}{\partial x}\\
\frac{\partial \hat{\mathbf{N}}_i}{\partial y}\\
\frac{\partial \hat{\mathbf{N}}_i}{\partial z}
\end{array}\right\}=(DF_E)^{-T}\left\{\begin{array}{c}
\frac{\partial \hat{\mathbf{N}}_i}{\partial \hat{x}}\\
\frac{\partial \hat{\mathbf{N}}_i}{\partial \hat{y}}\\
\frac{\partial \hat{\mathbf{N}}_i}{\partial \hat{z}}
\end{array}\right\}
\end{equation*}
Denoting the determinant of $DF_E(\mathbf{\hat{r}})$ by $J_E$, the integrals leading upto the element stiffness matrix and force vectors are given as
\begin{equation}
\left.
\begin{aligned}
&\int\limits_{E}\mathbf{N}^T\mathbf{f}\equiv \int\limits_{\hat{E}}\hat{\mathbf{N}}^T\mathbf{f}J_E
\equiv \sum\limits_{g\in \mathscr{G}}(\hat{\mathbf{N}}^T\mathbf{f}J_E)\vert_{g}w_g\\
&\int\limits_{E} \mathbf{B}^T\alpha \bar{p} \mathbf{I} \equiv \int\limits_{\hat{E}} \hat{\mathbf{B}}^T\alpha \bar{p} \mathbf{I}J_E\,\equiv \sum\limits_{g\in \mathscr{G}}(\hat{\mathbf{B}}^T\alpha \bar{p} \mathbf{I}J_E)\vert_{g}w_g\\
&\int\limits_{E} \mathbf{B}^T \mathbb{D} \mathbf{B}\equiv \int\limits_{\hat{E}} \hat{\mathbf{B}}^T \mathbb{D} \hat{\mathbf{B}} J_E\equiv \sum\limits_{g\in \mathscr{G}}(\hat{\mathbf{B}}^T \mathbb{D} \hat{\mathbf{B}} J_E)\vert_{g}w_g \\
&\int\limits_{\partial E\cap \Gamma_N^p} \mathbf{N}^T\mathbf{t} \equiv \int\limits_{\partial \hat{E}\cap \Gamma_N^p} \hat{\mathbf{N}}^T\mathbf{t}\,J_E \equiv \sum\limits_{g'\in \mathscr{G}'}(\hat{\mathbf{N}}^T\mathbf{t}\,J_E)\vert_{g'} w_{g'}
\end{aligned}
\right\}
\label{integrals}
\end{equation}
where the set $\mathscr{G}$ of quadrature points with associated weights is
\begin{align*}
&g_1\equiv \big(\frac{1}{2}\big(1-\frac{\sqrt{3}}{3}\big),\frac{1}{2}\big(1-\frac{\sqrt{3}}{3}\big),\frac{1}{2}\big(1-\frac{\sqrt{3}}{3}\big)\big)\\
&g_2\equiv \big(\frac{1}{2}\big(1+\frac{\sqrt{3}}{3}\big),\frac{1}{2}\big(1-\frac{\sqrt{3}}{3}\big),\frac{1}{2}\big(1-\frac{\sqrt{3}}{3}\big)\big)\\
&g_3\equiv \big(\frac{1}{2}\big(1+\frac{\sqrt{3}}{3}\big),\frac{1}{2}\big(1+\frac{\sqrt{3}}{3}\big),\frac{1}{2}\big(1-\frac{\sqrt{3}}{3}\big)\big)\\
&g_4\equiv \big(\frac{1}{2}\big(1-\frac{\sqrt{3}}{3}\big),\frac{1}{2}\big(1+\frac{\sqrt{3}}{3}\big),\frac{1}{2}\big(1-\frac{\sqrt{3}}{3}\big)\big)\\
&g_5\equiv \big(\frac{1}{2}\big(1-\frac{\sqrt{3}}{3}\big),\frac{1}{2}\big(1-\frac{\sqrt{3}}{3}\big),\frac{1}{2}\big(1+\frac{\sqrt{3}}{3}\big)\big)\\
&g_6\equiv \big(\frac{1}{2}\big(1+\frac{\sqrt{3}}{3}\big),\frac{1}{2}\big(1-\frac{\sqrt{3}}{3}\big),\frac{1}{2}\big(1+\frac{\sqrt{3}}{3}\big)\big)\\
&g_7\equiv \big(\frac{1}{2}\big(1+\frac{\sqrt{3}}{3}\big),\frac{1}{2}\big(1+\frac{\sqrt{3}}{3}\big),\frac{1}{2}\big(1+\frac{\sqrt{3}}{3}\big)\big)\\
&g_8\equiv \big(\frac{1}{2}\big(1-\frac{\sqrt{3}}{3}\big),\frac{1}{2}\big(1+\frac{\sqrt{3}}{3}\big),\frac{1}{2}\big(1+\frac{\sqrt{3}}{3}\big)\big)\\
&w_g=\frac{1}{8}\qquad \forall\,\,g\in \mathscr{G}
\end{align*}
and the set $\mathscr{G}'$ of quadrature points with associated weights is
\begin{align*}
&g'_1\equiv \big(\frac{1}{2}\big(1-\frac{\sqrt{3}}{3}\big),\frac{1}{2}\big(1-\frac{\sqrt{3}}{3}\big)\big)\\
&g'_2\equiv \big(\frac{1}{2}\big(1+\frac{\sqrt{3}}{3}\big),\frac{1}{2}\big(1-\frac{\sqrt{3}}{3}\big)\big)\\
&g'_3\equiv \big(\frac{1}{2}\big(1+\frac{\sqrt{3}}{3}\big),\frac{1}{2}\big(1+\frac{\sqrt{3}}{3}\big)\big)\\
&g'_4\equiv \big(\frac{1}{2}\big(1-\frac{\sqrt{3}}{3}\big),\frac{1}{2}\big(1+\frac{\sqrt{3}}{3}\big)\big)\\
&w_{g'}=\frac{1}{4}\qquad \forall\,\,g'\in \mathscr{G}'
\end{align*}
In lieu of \eqref{poroeq1} and \eqref{integrals}, we write
\begin{equation}
\left.\begin{array}{c} 
\label{poroeq2}
\mathbf{K}\mathcal{U}=\mathbf{F}\\
\mathbf{K}=\sum\limits_{E\in \mathscr{T}_h}\sum\limits_{g\in \mathscr{G}}(\hat{\mathbf{B}}^T \mathbb{D} \hat{\mathbf{B}} J_E)\vert_{g}w_g \\
\mathbf{F}=
\sum\limits_{E\in \mathscr{T}_h}[\sum\limits_{g\in \mathscr{G}}(\hat{\mathbf{B}}^T\alpha \bar{p} \mathbf{I}J_E)\vert_{g}w_g+\sum\limits_{g\in \mathscr{G}}(\hat{\mathbf{N}}^T\mathbf{f}J_E)\vert_{g}w_g+\sum\limits_{g'\in \mathscr{G}'}(\hat{\mathbf{N}}^T\mathbf{t}\,J_E)\vert_{g'} w_{g'}]
\end{array}\right\}
\end{equation}
\section{Conclusions and outlook}
The model equations and algorithmic framework for immiscible two phase flow coupled with linear poromechanics have been presented. The nonlinear nature of the equations of two phase flow neccesitate the need for nonlinear iterations in the manner of the Newton-Raphson method. The fixed mean stress constraint is imposed on the system of equations in each Newton iteration. The pressure solution acts as a forcing term for the poromechanical system. The displacement solution is then post-processed to obtain volumetric strains. These strains are then updated and act as poromechanical feedback to the two phase flow system.
\appendix
\section{Parabolic elliptic nature of the differential equation of single phase flow}\label{dis1}
The equation of continuity for single phase flow along with the Darcy's law are given by
\begin{align}
\label{one1}
&\frac{\partial (\phi\rho)}{\partial t}+\nabla \cdot (\rho\mathbf{v})=q\\
\label{two1}
&\mathbf{v}=-\mathbf{K}\rho(\nabla p-\mathbf{g})
\end{align}
\eqref{one1} can also be written as
\begin{align}
\label{three1}
&\phi\frac{1}{\rho}\frac{\partial \rho}{\partial t}+\frac{\partial \phi}{\partial t}+\mathbf{v}\cdot \frac{1}{\rho}\nabla \rho+\nabla \cdot \mathbf{v}=\frac{q}{\rho}
\end{align}
With the fluid being slightly compressible with compressibility $c>0$, we get the following 
\begin{align*}
&\frac{\partial \rho}{\partial t}=c\rho\frac{\partial p}{\partial t}\\
&\nabla \rho=c\rho\nabla p
\end{align*}
as a result of which \eqref{three1} is written as
\begin{align}
\nonumber
&\phi c\frac{\partial p}{\partial t}+\frac{\partial \phi}{\partial t}+\mathbf{v}\cdot c\nabla p+\nabla \cdot \mathbf{v}=\frac{q}{\rho}
\end{align}
Substituting \eqref{two1} in the above, we get
\begin{align}
\label{four1}
&\phi c\frac{\partial p}{\partial t}+\frac{\partial \phi}{\partial t}-(\mathbf{K}\rho(\nabla p-\mathbf{g}))\cdot c\nabla p-\nabla \cdot (\mathbf{K}\rho(\nabla p-\mathbf{g}))=\frac{q}{\rho}
\end{align}
\eqref{four1} is parabolic in $p$ and degenerates to an elliptic equation in $p$ in case the fluid is incompressible ($c=0$) as follows
\begin{align}
\label{five1}
&\frac{\partial \phi}{\partial t}-\nabla \cdot (\mathbf{K}\rho(\nabla p-\mathbf{g}))=\frac{q}{\rho}
\end{align}
\section{Parabolic hyperbolic elliptic nature of the differential equations of immiscible two phase flow}\label{dis2}
The key to the understanding of the nature of the two phase equations lies in their reduction to a system of one parabolic saturation equation conventionally posed in terms of the saturation of the wetting phase (water in water-oil systems) coupled with a parabolic pressure equation posed in terms of a global, or the reduced, or the intermediate pressure~\cite{chavent}. 
\begin{itemize}
\item The parabolic pressure equation degenerates to an elliptic equation in case both the fluids are incompressible.
\item The parabolic saturation equation degenerates to a first order hyperbolic equation in case the wetting fluid is incompressible and the capillary pressure is ignored.
\end{itemize}
The equations of continuity for immiscible two phase flow along with the Darcy's law are given by
\begin{align}
\label{one}
&\frac{\partial (\phi\rho_{\beta}S_{\beta})}{\partial t}+\nabla \cdot (\rho_{\beta}\mathbf{v}_{\beta})=q_{\beta}\\
\label{two}
&\mathbf{v}_{\beta}=-\mathbf{K}\chi_{\beta}(\nabla p_{\beta}-\rho_{\beta} \mathbf{g})
\end{align}
where $\beta\equiv w$ for the water (wetting) phase, $\beta\equiv o$ for the oil (non-wetting) phase, $\mathbf{v}_{\beta}$ is the volumetric velocity of the phase $\beta$ and $\chi_{\beta}\equiv \frac{k_{r\beta}}{\mu_{\beta}}$. The equations \eqref{one} can also be written as
\begin{align}
\label{three}
&S_{\beta}\frac{\partial \phi}{\partial t}+\phi\frac{\partial S_{\beta}}{\partial t}+\frac{\phi S_{\beta}}{\rho_{\beta}}\frac{\partial \rho_{\beta}}{\partial t}+\nabla \cdot \mathbf{v}_{\beta}+\mathbf{v}_{\beta}\cdot \frac{\nabla \rho_{\beta}}{\rho_{\beta}}=\frac{q_{\beta}}{\rho_{\beta}}
\end{align}
With both fluids being slightly compressible with compressibilities $c_{\beta}>0$, we get the following 
\begin{align*}
&\frac{\partial \rho_{\beta}}{\partial t}=c_{\beta}\rho_{\beta}\frac{\partial p_{\beta}}{\partial t}\\
&\nabla \rho_{\beta}=c_{\beta}\rho_{\beta}\nabla p_{\beta}
\end{align*}
as a result of which \eqref{three} can be written as
\begin{align}
\label{four}
&S_{\beta}\frac{\partial \phi}{\partial t}+\phi\frac{\partial S_{\beta}}{\partial t}+\phi S_{\beta}c_{\beta}\frac{\partial p_{\beta}}{\partial t}+\nabla \cdot \mathbf{v}_{\beta}+\mathbf{v}_{\beta}\cdot c_{\beta}\nabla p_{\beta}=\frac{q_{\beta}}{\rho_{\beta}}
\end{align}
Adding the individual phase equations \eqref{four} and invoking the relation $S_o+S_w=1$ results in
\begin{align*}
&\frac{\partial \phi}{\partial t}+\frac{\phi (1-S_w)}{\rho_{o}}\frac{\partial \rho_{o}}{\partial t}+c_o\phi S_o\frac{\partial p_o}{\partial t}
+c_w\phi S_w\frac{\partial p_w}{\partial t}
+\mathbf{v}_o\cdot c_o \nabla p_o
+\mathbf{v}_w\cdot c_w \nabla p_w\\
&+\nabla \cdot(\mathbf{v}_o+\mathbf{v}_w)=\frac{q_o}{\rho_o}+\frac{q_w}{\rho_w}
\end{align*}
Introducing a `total velocity' $\mathbf{v}\equiv \mathbf{v}_o+\mathbf{v}_w$ in the above equation results in
\begin{align}
\label{pressureq}
&\nabla \cdot \mathbf{v}=-\frac{\partial \phi}{\partial t}
-c_o\phi S_o\frac{\partial p_o}{\partial t}
-c_w\phi S_w\frac{\partial p_w}{\partial t}
-\mathbf{v}_o\cdot c_o \nabla p_o
-\mathbf{v}_w\cdot c_w \nabla p_w
+\frac{q_o}{\rho_o}+\frac{q_w}{\rho_w}
\end{align}
The total velocity is given by
\begin{align*}
\mathbf{v}=\mathbf{v}_o+\mathbf{v}_w=-\mathbf{K}\chi_{o}(\nabla p_{o}-\rho_{o} \mathbf{g})-\mathbf{K}\chi_{w}(\nabla p_{w}-\rho_{w} \mathbf{g})
\end{align*}
Invoking the capillary pressure $p_c=p_0-p_w$ in the above, we get
\begin{align*}
&\mathbf{v}=\mathbf{v}_o+\mathbf{v}_w=-\mathbf{K}\chi_{o}(\nabla p_{o}-\rho_{o} \mathbf{g})-\mathbf{K}\chi_{w}(\nabla p_{o}-\nabla p_{c}-\rho_{w} \mathbf{g})\\
&=-\mathbf{K}(\chi_{o}+\chi_{w})\nabla p_{o}+\mathbf{K}\chi_{w}\nabla p_{c}+\mathbf{K}\chi_{o}\rho_{o} \mathbf{g}+\mathbf{K}\chi_{w}\rho_{w} \mathbf{g}
\end{align*}
Introducing $\chi=\chi_o+\chi_w$ in the above equation, we get
\begin{align*}
&\mathbf{v}=-\mathbf{K}\chi\nabla p_{o}+\mathbf{K}\chi_{w}\nabla p_{c}+\mathbf{K}\chi_{o}\rho_{o} \mathbf{g}+\mathbf{K}\chi_{w}\rho_{w} \mathbf{g}\\
&=-\mathbf{K}\chi(\nabla p_{o}-\frac{\chi_w}{
\chi}\nabla p_{c}-\frac{\chi_o}{
\chi}\rho_{o} \mathbf{g}-\frac{\chi_w}{
\chi}\rho_{w} \mathbf{g})
\end{align*}
Introducing $G=\frac{\chi_o}{
\chi}\rho_{o} \mathbf{g}+\frac{\chi_w}{
\chi}\rho_{w} \mathbf{g}$ in the above equation, we get
\begin{align}
\label{totalinterim}
&\mathbf{v}=-\mathbf{K}\chi(\nabla p_{o}-\frac{\chi_w}{
\chi}\nabla p_{c}-G)
\end{align}
Further, introducing a `global pressure' $p$ such that
\begin{align}
\label{global}
\nabla p\equiv \nabla p_{o}-\frac{\chi_w}{
\chi}\nabla p_{c}
\end{align}
we get the following
\begin{align}
\label{total}
&\mathbf{v}=-\mathbf{K}\chi(\nabla p-G)
\end{align}
\subsection{The pressure equation}
Substituting \eqref{total} in \eqref{pressureq}, we get
\begin{align}
\nonumber
&-\nabla \cdot (\mathbf{K}\chi(\nabla p-G))\\
\label{pressureq2}
&=-\frac{\partial \phi}{\partial t}
-c_o\phi S_o\frac{\partial p_o}{\partial t}
-c_w\phi S_w\frac{\partial p_w}{\partial t}
-\mathbf{v}_o\cdot c_o \nabla p_o
-\mathbf{v}_w\cdot c_w \nabla p_w
+\frac{q_o}{\rho_o}+\frac{q_w}{\rho_w}
\end{align}
\eqref{pressureq2} is parabolic in global pressure $p$ and degenerates to an elliptic equation in $p$ in case both the fluids are incompressible ($c_{\beta}=0$) as follows
\begin{align}
\nonumber
&-\nabla \cdot (\mathbf{K}\chi(\nabla p-G))=-\frac{\partial \phi}{\partial t}
+\frac{q_o}{\rho_o}+\frac{q_w}{\rho_w}
\end{align}
We rewrite \eqref{global} as
\begin{align*}
&\nabla p\equiv \nabla p_{o}-\frac{\chi_w}{
\chi}\nabla p_{c}=\frac{1}{2}(\nabla p_{o}+\nabla p_{w})+\frac{1}{2}(\nabla p_{o}-\nabla p_{w})-\frac{\chi_w}{\chi}\nabla p_{c}\\
&=\frac{1}{2}(\nabla p_{o}+\nabla p_{w})+\frac{1}{2}\nabla p_c-\frac{\chi_w}{\chi}\nabla p_{c}\\
&=\frac{1}{2}(\nabla p_{o}+\nabla p_{w})+\frac{1}{2}\big(\frac{\chi_o-\chi_w}{\chi}\big)\nabla p_c
\end{align*}
Integrating the above equation results in an expression for the global pressure as follows
\begin{align*}
p=\frac{1}{2}(p_{o}+p_{w})+\frac{1}{2}\int\limits_{S_c}^{S_w}\big(\frac{\chi_o-\chi_w}{\chi}\big)\frac{\partial p_c}{\partial S_w}\nabla S_w
\end{align*}
where $S_c$ is the value of water saturation at which the capillary pressure is zero.
\subsection{The saturation equation}
In lieu of \eqref{four}, the equation of continuity of the water phase is
\begin{align}
\label{five}
&S_{w}\frac{\partial \phi}{\partial t}+\phi\frac{\partial S_{w}}{\partial t}+\phi S_{w}c_{w}\frac{\partial p_{w}}{\partial t}+\nabla \cdot \mathbf{v}_{w}+\mathbf{v}_{w}\cdot c_{w}\nabla p_{w}=\frac{q_{w}}{\rho_{w}}
\end{align}
Now, multiplying \eqref{totalinterim} by $\frac{\chi_w}{\chi}$, we get
\begin{align*}
\frac{\chi_w}{\chi}\mathbf{v}=-\mathbf{K}\chi_w(\nabla p_{o}-\frac{\chi_w}{
\chi}\nabla p_{c}-G)
\end{align*}
Substituting $p_o=p_c+p_w$ in the above equation results in
\begin{align*}
&\frac{\chi_w}{\chi}\mathbf{v}=-\mathbf{K}\chi_w(\nabla p_{c}+\nabla p_w-\frac{\chi_w}{
\chi}\nabla p_{c}-G)\\
&=-\mathbf{K}\chi_w(\nabla p_w-\rho_w\mathbf{g}+\frac{\chi_o}{
\chi}\nabla p_{c}-G+\rho_w\mathbf{g})\\
&=-\mathbf{K}\chi_w(\nabla p_w-\rho_w\mathbf{g})
-\mathbf{K}\chi_w(\frac{\chi_o}{
\chi}\nabla p_{c}-\frac{\chi_o}{
\chi}\rho_{o} \mathbf{g}-\frac{\chi_w}{
\chi}\rho_{w} \mathbf{g}+\rho_w\mathbf{g})\\
&=\mathbf{v}_w-\mathbf{K}\chi_w\frac{\chi_o}{
\chi}(\nabla p_{c}-(\rho_{o}-\rho_{w})\mathbf{g})
\end{align*}
which can be written as
\begin{align}
\label{six}
\mathbf{v}_w=\frac{\chi_w}{\chi}\mathbf{v}+\mathbf{K}\chi_w\frac{\chi_o}{
\chi}(\nabla p_{c}-(\rho_{o}-\rho_{w})\mathbf{g})
\end{align}
Substituting \eqref{six} in \eqref{five}, we get
\begin{align}
\nonumber
&S_{w}\frac{\partial \phi}{\partial t}+\phi\frac{\partial S_{w}}{\partial t}+\phi S_{w}c_{w}\frac{\partial p_{w}}{\partial t}+\nabla \cdot (\frac{\chi_w}{\chi}\mathbf{v}+\mathbf{K}\chi_w\frac{\chi_o}{
\chi}(\nabla p_{c}-(\rho_{o}-\rho_{w})\mathbf{g}))\\
\label{seven}
&+(\frac{\chi_w}{\chi}\mathbf{v}+\mathbf{K}\chi_w\frac{\chi_o}{
\chi}(\nabla p_{c}-(\rho_{o}-\rho_{w})\mathbf{g}))\cdot c_{w}\nabla p_{w}=\frac{q_{w}}{\rho_{w}}
\end{align}
\eqref{seven} is parabolic in water saturation $S_w$ and degenerates to a first order hyperbolic equation in $S_w$ in case water is incompressible ($c_w=0$) and capillary pressure is ignored ($p_c=0$) as follows
\begin{align}
\label{eight}
&S_{w}\frac{\partial \phi}{\partial t}+\phi\frac{\partial S_{w}}{\partial t}+\nabla \cdot (\frac{\chi_w}{\chi}\mathbf{v}-\mathbf{K}\chi_w\frac{\chi_o}{
\chi}(\rho_{o}-\rho_{w})\mathbf{g})=\frac{q_{w}}{\rho_{w}}
\end{align}
\section{Mixed finite element space for flux}\label{bddfa}
\begin{figure}[h]
\centering
\includegraphics[scale=0.5]{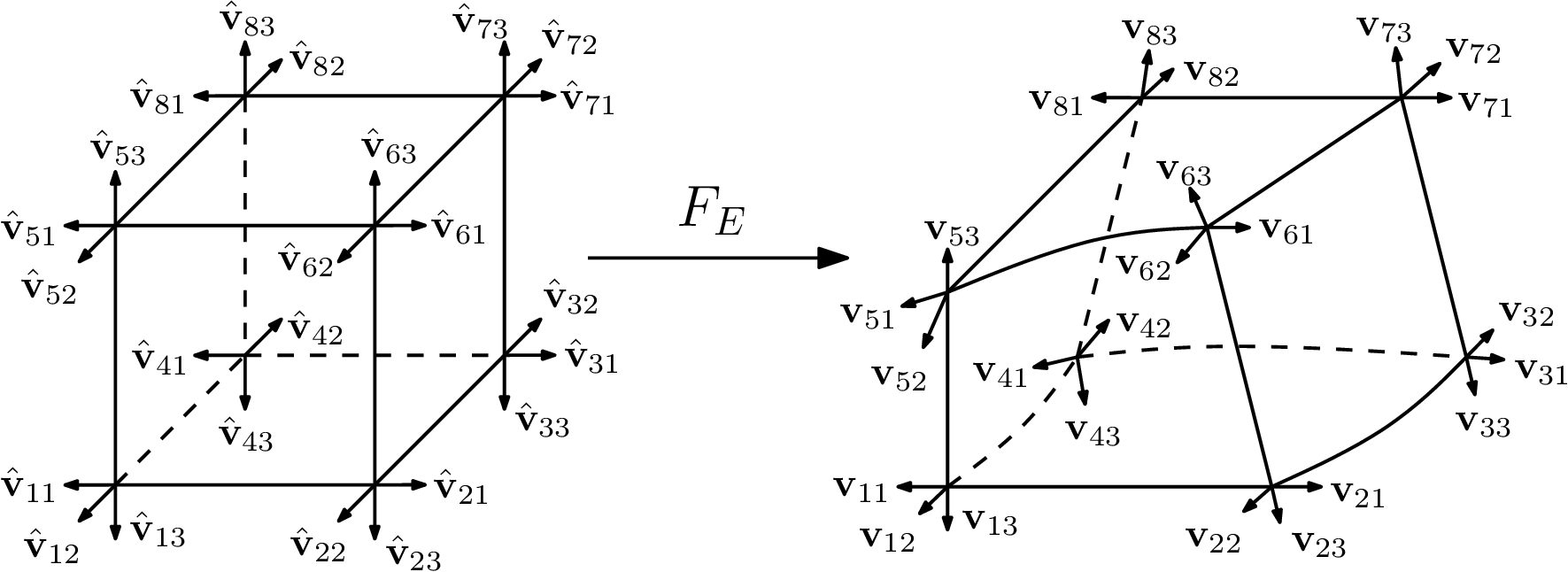}
\caption{Degrees of freedom and basis functions for the enhanced BDDF$_1$ velocity space on hexahedra.}
\label{bddf}
\end{figure}
Let $\mathbf{V}^*_h\times W_h$ be the lowest order BDDF$_1$ MFE spaces on hexahedra~\cite{ref17}. With $\mathbf{x}\equiv(\hat{x},\hat{y},\hat{z})\in \hat{E}$, these spaces are defined on $\hat{E}$ as
\begin{align*}
\hat{\mathbf{V}}^*(\hat{E})&=(\mathbb{P}_1(\hat{E}))^3
+r_0\,curl(0,0,\hat{x}\hat{y}\hat{z})^T
+r_1\,curl(0,0,\hat{x}\hat{y}^2)^T
+s_0\,curl(\hat{x}\hat{y}\hat{z},0,0)^T\\
&+s_1\,curl(\hat{y}\hat{z}^2,0,0)^T
+t_0\,curl(0,\hat{x}\hat{y}\hat{z},0)^T
+t_1\,curl(0,\hat{x}^2\hat{z},0)^T\\
\hat{W}(\hat{E})&=\mathbb{P}_0(\hat{E})
\end{align*}
with the following properties
\begin{align*}
\hat{\nabla} \cdot \hat{\mathbf{V}}^*(\hat{E})=\hat{W}(\hat{E}), \qquad \mathrm{and} \qquad \forall \hat{\mathbf{v}}\in \hat{\mathbf{V}}^*(\hat{E}),\,\,\forall \hat{e}\subset \partial \hat{E},\,\,\hat{\mathbf{v}}\cdot \hat{\mathbf{n}}_{\hat{e}}\in \mathbb{P}_1(\hat{e})
\end{align*}
where $\hat{e}$ represents a face of $\hat{E}$ and $\hat{\mathbf{n}}_{\hat{e}}$ the unit outward normal to $\hat{e}$. The multipoint flux approximation procedure requires on each face one velocity degree of freedom to be associated with each vertex thus requiring four degrees of freedom per face. Since $\mathbf{V}^*_h$ has only three degrees of freedom per face, it is augmented with one degree of freedom per face resulting in addition of six degrees of freedom per element. Since the properties of constant divergence, linear independence of the shape functions and continuity of the normal component across the element faces are to be preserved, six curl terms are added~\cite{ingram} to $\mathbf{V}^*_h$. Let $\mathbf{V}_h\times W_h$ be the enhanced BDDF$_1$ spaces on hexahedra. On $\hat{E}$, these spaces are
\begin{align*}
\hat{\mathbf{V}}(\hat{E})&=\hat{\mathbf{V}}^*(\hat{E})
+r_2\,curl(0,0,\hat{x}^2\hat{z})^T
+r_3\,curl(0,0,\hat{x}^2\hat{y}\hat{z})^T
+s_2\,curl(\hat{x}\hat{y}^2,0,0)^T\\
&+s_3\,curl(\hat{x}\hat{y}^2\hat{z}^2,0,0)^T
+t_2\,curl(0,\hat{y}\hat{z}^2,0)^T
+t_3\,curl(0,\hat{x}\hat{y}\hat{z}^2,0)^T\\
\hat{W}(\hat{E})&=\mathbb{P}_0(\hat{E})
\end{align*}
with the following properties
\begin{align*}
\hat{\nabla} \cdot \hat{\mathbf{V}}(\hat{E})=\hat{W}(\hat{E}), \qquad \mathrm{and} \qquad \forall \hat{\mathbf{v}}\in \hat{\mathbf{V}}(\hat{E}),\,\,\forall \hat{e}\subset \partial \hat{E},\,\,\hat{\mathbf{v}}\cdot \hat{\mathbf{n}}_{\hat{e}}\in \mathbb{Q}_1(\hat{e})
\end{align*}
Since $dim\,\mathbb{Q}_1(\hat{e})=4$, the dimension of $\hat{\mathbf{V}}(\hat{E})$ is 24 as shown in Figure \ref{bddf}.
\bibliographystyle{unsrt} 
\bibliography{diss}
\end{document}